\documentclass[10pt]{IEEEtran}

\usepackage{cite}

\usepackage{multirow}
\usepackage{subfigure} 
\usepackage{graphicx}
\usepackage[safe]{tipa}
\usepackage{xcolor}
\usepackage{amsmath}
\usepackage{amssymb}
\usepackage{setspace}

\renewcommand{\baselinestretch}{1.2}

\begin{document}
\title{Temporarily-Aware Context Modelling using Generative Adversarial Networks for Speech Activity Detection}

\author{Tharindu~Fernando,~\IEEEmembership{Student Member,~IEEE,}
	Sridha~Sridharan,~\IEEEmembership{Life Senior Member,~IEEE,}
	Mitchell~McLaren,
	Darshana~Priyasad,~\IEEEmembership{Member,~IEEE,}
        Simon~Denman,~\IEEEmembership{Member,~IEEE,}
        ~and~ Clinton~Fookes,~\IEEEmembership{Senior Member,~IEEE.}
         
         \IEEEcompsocitemizethanks{\IEEEcompsocthanksitem T. Fernando, S. Sridharan, D. Priyasad, S. Denman, C. Fookes  are with Speech Research Lab, SAIVT, Queensland University of Technology, Australia. M. McLaren is with Speech Technology and Research Laboratory of SRI International.\protect\\

E-mail: t.warnakulasuriya@qut.edu.au
}\thanks{Manuscript received }}

\markboth{IEEE Transactions on Audio, Speech, and Language Processing,~Vol.~14, No.~8, August~2015}%
{Fernando \MakeLowercase{\textit{et al.}}: Time Aware Generative Adversarial Networks for Speech Activity Detection}
\maketitle

\begin{abstract}
This paper presents a novel framework for Speech Activity Detection (SAD). Inspired by the recent success of multi-task learning approaches in the speech processing domain, we propose a novel joint learning framework for SAD. We utilise generative adversarial networks to automatically learn a loss function for joint prediction of the frame-wise speech/ non-speech classifications together with the next audio segment. In order to exploit the temporal relationships within the input signal, we propose a temporal discriminator which aims to ensure that the predicted signal is temporally consistent. We evaluate the proposed framework on multiple public benchmarks, including NIST OpenSAT' 17, AMI Meeting and HAVIC, where we demonstrate its capability to outperform state-of-the-art SAD approaches. Furthermore, our cross-database evaluations demonstrate the robustness of the proposed approach across different languages, accents, and acoustic environments.
\end{abstract}


\IEEEpeerreviewmaketitle

\section{Introduction}
\IEEEPARstart{S}{peech} Activity Detection (SAD) plays a pivotal role in many speech processing systems. Despite the consistent progress attained in this subject, the problem is far from being solved as evidenced by evaluation results across the vast variety of acoustic conditions featured in challenging benchmarks such as HAVIC \cite{havic_ldc} and NIST OpenSAT' 17 \cite{nist_open_SAT}. 

Our work is inspired by recent observations in speech processing where multi-task learning approaches have shown to outperform single task learning methods in numerous areas, including, speech synthesis \cite{chen2015multi}, speech recognition \cite{pironkov2016multi}, speech enhancement \cite{chen2015speech}, and speech emotion recognition \cite{kim2017speech}. For instance, the seminal work by Pironkov et. al  \cite{pironkov2016multi} demonstrated that significant improvements in the accuracy of Automatic Speech Recognition (ASR) can be obtained by combining the ASR task with context recognition and gender classification as auxiliary tasks, as opposed to performing ASR alone. Furthermore, the evaluations in \cite{chen2015speech,kim2017speech} suggested that methods learned using the multi-task learning paradigm are not only robust when evaluated in cross database scenarios, but also learn powerful and more discriminative features to facilitate both tasks. 

Inspired by these findings, we exploit the power of Generative Adversarial Networks (GAN) \cite{isola2017image,pascual2017segan} to accurately perform speech/non-speech classification together with an auxiliary task. In choosing the appropriate auxiliary task for SAD we draw inspiration from a conclusion in the field of neuroscience that humans recognise speech in noisy conditions through the awareness of the next segment of speech which is most likely to be heard \cite{leonard2015dynamic,heinrich2008illusory}. We therefore chose the prediction of the next audio segment as the auxiliary task as it also complements the primary SAD task via learning the context of the input audio embedding. Through the prediction of next audio segment our model tries to learn a contextual mapping between the input audio segments and the next segment which is likely to be heard. 

Even though the final speech activity decision is agnostic to the actual content of speech, there are reasons to conjecture that the SAD accuracy could be improved by making use the semantic information of speech. It is known that humans make use the semantic information to understand speech that is affected significantly by noise \cite{leonard2015dynamic,heinrich2008illusory}. In \cite{darwin2008listening} the authors demonstrated that our inferior-frontal cortex predicts what someone is likely to hear next even before the actual sound reaches the superior temporal gyrus, allowing us to separate noise from what is actually spoken. One of our aims in this paper is to investigate how and to what extent we could improve the performance of SAD if we were to use semantic information to predict the next speech segment. Current SAD methods simply classify whether a sample is speech or non-speech, without paying attention to the temporal context. Even though the current state-of-the-art SAD systems extract features from a sliding window surrounding the event frame of interest, they consider the frame as an isolated event and do not consider the entire sequence when detecting the speech activity. We show in this paper that through the prediction of the next audio segment by exploiting the task-specific loss-function learning capability of the GAN framework, we can improve SAD accuracy by a significant amount.

\begin{figure*}[htbp]
  \centering
       \includegraphics[width=.9\linewidth]{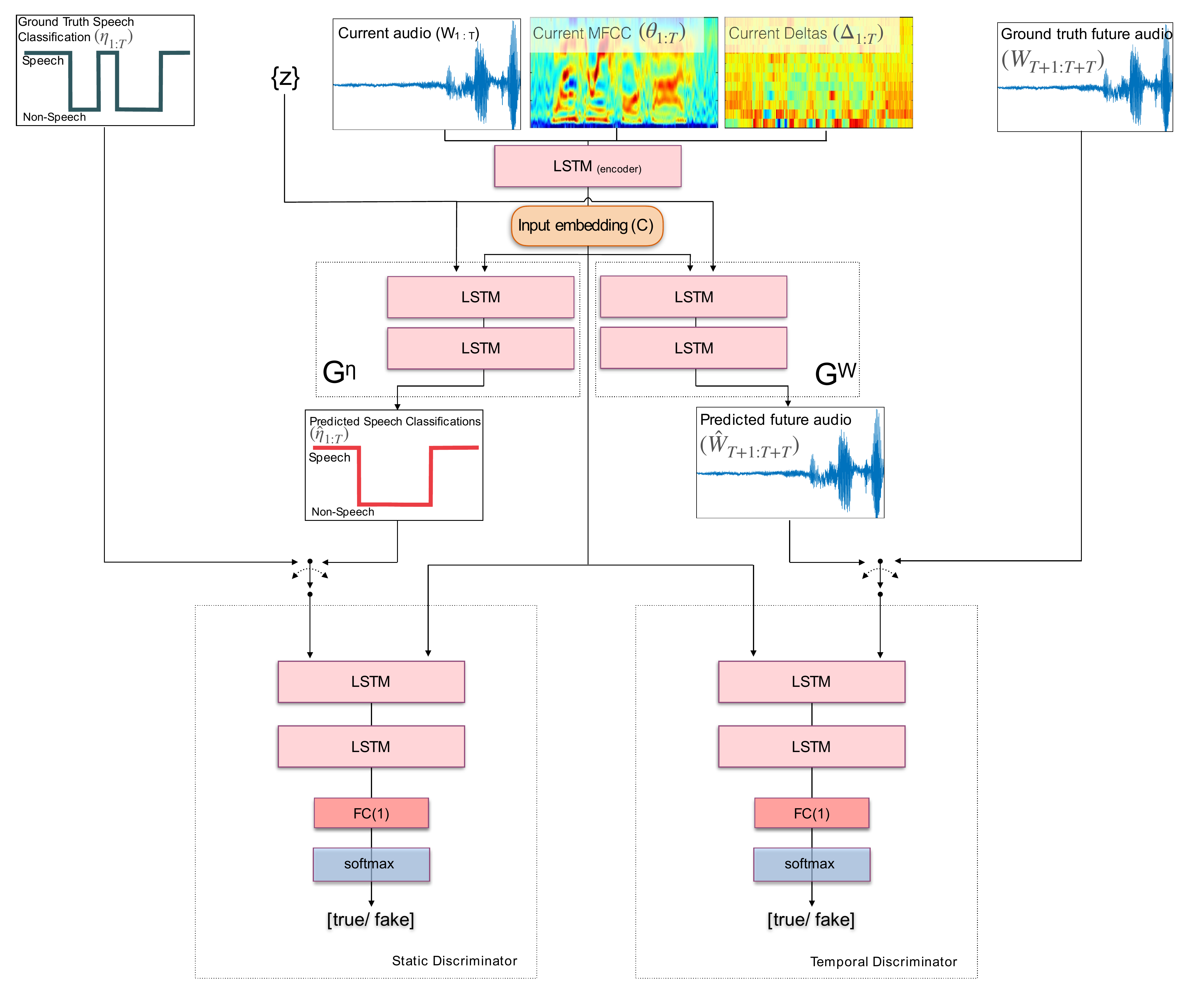}
   \caption{Proposed TA-GAN framework: Given the current time $\tau$, the model input is a segment containing the T audio frames and the features extracted from this segment (where $w_t$ is the raw audio of frame $t$, $\theta_t$ denotes the MFCC feature \cite{mccowan2011delta} and $\Delta_t$ denotes the MFCC deltas \cite{zhang2018speech} for the same frame) directly proceeding $\tau$ and we term this the \textit{current segment}. The encoder receives audio, MFCC and Deltas of MFCC inputs and embeds this information in an input embedding, $C$. Using this embedding and a random noise vector $z$, the classification sequence generator, $G^{\eta}$, synthesises a frame-wise speech classification sequence for the current time window while the same $C$ and a random noise vector $z$ are used by the audio generator, $G^{w}$, to synthesise the audio signal for the next T frames directly following $\tau$ which we term the \textit{future segment}. We utilise two discriminators. The static discriminator, $D^{\eta}$, receives the current input embeddings and either the synthesised or ground truth speech classification sequences and tries to discriminate between the two. The temporal discriminator, $D^{w}$, receives the current input embeddings and either the synthesised or ground truth future audio segments and learns to classify them, considering the temporal relationships between audio frames within those signals.}
  \label{fig:TA_GAN_model}
\end{figure*}

The proposed architecture is shown in Fig. \ref{fig:TA_GAN_model}. The model utilises audio, Mel-Frequency Cepstral Coefficients (MFCC) and Deltas of MFCC as the inputs and encodes these inputs into an encoded representation, $C$. The generators receive this input embedding, $C$, and a noise vector, $z$, as the inputs. We utilise two generators, $G^{\eta}$, for synthesising the frame-wise speech/ non-speech classifications and $G^{w}$, which synthesises the audio signal for the next time window. It should be noted that in Fig. \ref{fig:TA_GAN_model} the generators, $G^{\eta}$ and $G^{w}$ are denoted as two separate LSTM blocks, each with two cells of LSTMs.The static discriminator, $D^{\eta}$, receives the current input embeddings and either the synthesised or ground truth speech classification sequences and tries to discriminate between the two. The temporal discriminator, $D^{w}$, also receives the current input embeddings and either the synthesised or ground truth future audio segments and learns to classify them, considering the temporal consistency of those signals.

The main contributions of the proposed work are summarised as follows: \begin{itemize}
\item We introduce a Temporarily-Aware GAN (TA-GAN) learning framework for speech activity detection.
\item We demonstrate how a custom loss function for speech activity detection can be automatically learned through the GAN learning process. 
\item We propose a novel temporal discriminator which encourages the generator to synthesise future speech segments in accordance with the current context. 
\item We perform extensive evaluations on the proposed framework using multiple public benchmarks and demonstrate performance beyond that of current state-of-the-art systems. 
\end{itemize}

\section{Related Work  on Supervised Speech Activity Detection }
In supervised SAD, machine learning algorithms are trained on annotated audio data to discriminate speech from non-speech segments. 
Several prior works have focused on finding better discriminative features for supervised classification \cite{ng2012developing,saon2013ibm,thomas2015improvements,graciarena2013all, mak2014study, shin2010voice,kumar2016voice}. For instance in \cite{graciarena2013all} the authors suggest a combination of  MFCCs and Gabor features. In \cite{drugman2016voice} the authors suggest the use of source and filter based features and perform a score level fusion. In \cite{ferrer2016phonetically} the authors propose the use of bottleneck features for predicting the speech and non-speech posteriors. In \cite{kinnunen2016happy} the authors fuse six SAD systems, two supervised and four unsupervised, for the NIST-Open-SAD-2015 challenge. Supervised systems utilise labelled speech and non-speech segments for training the SAD while the unsupervised methods utilise a fixed or adaptive threshold for the SAD task. The work of Hwang et. al \cite{hwang2016ensemble} proposed the utilisation of an ensemble of deep neural networks trained on different noise types for supervised SAD. In a different line of work, \cite{sholokhov2018semi} proposed a semi-supervised learning approach for GMM training, using power normalized cepstral coefficients, perceptual linear prediction coefficients, and frequency domain linear prediction as features in addition to MFCCs. 

However, none of the above stated deep learning systems have explicitly modelled the temporal relationship between audio frames in the input signal when performing SAD.

One of the earliest attempts to leverage temporal modelling in SAD was based on Recurrent Neural Networks (RNNs) \cite{hughes2013recurrent} where the authors demonstrate a reduction of 26\% in the false alarm rate compared to their Gaussian Mixture Model (GMM) baseline which doesn't use any temporal modelling. In \cite{gelly2018optimization} the authors build upon this work where they augment the Long Short Term Memory (LSTM) cell architecture. They propose a coordinated-gate LSTM structure and a methodology to directly optimise the SAD loss using the Frame Error Rate (FER). Most recently, the Adaptive Context Attention Model (ACAM) \cite{kim2018voice} model extended the LSTM based temporal modelling scheme using an attention strategy to learn the context of the speech signal for noise robustness in the SAD system. In a different line of work, an audiovisual SAD system is proposed in \cite{tao2019end} in order to improve the robustness of the framework. 

\section{The Proposed Approach}
We are inspired by the tremendous success of DNN based multi-task learning frameworks \cite{chen2015speech, kim2017speech, pironkov2016multi} in speech processing which demonstrate greater robustness compared to single task learning methods. Motivated by these findings we investigate the utility of multi-task learning for SAD. To the best of our knowledge, the work in this study is the first to consider multi-task learning for SAD. Specifically, we attain joint predictions of the frame-wise speech/ non-speech classification along with the next audio segment through the proposed multi-task learning framework. As there doesn't exist an optimal, off the shelf loss function for the joint task that we are attaining, we utilise the GAN learning framework to automatically learn a loss function for these tasks. 

We exploit the task-specific loss-function learning capability of the GAN framework to automatically learn a custom loss function \cite{fernando2019memory,wang2018application,larsen2015autoencoding,gammulle2018multi} that facilitates these two tasks . The merit of this approach is that it allows us to learn a highly non-linear loss, in contrast to a linear loss like cross entropy, to optimally capture the underlying semantics of the process. This custom loss function learning capability of GANs is highly beneficial in the multi-task learning setting, as it allows us to learn a custom loss function that accounts for all the tasks at hand rather than simply adding together the loss functions for individual tasks. For instance, in \cite{gammulle2018multi} the authors illustrate the utility of GANs for video based action prediction while synthesising future frame representations, and the authors in \cite{he2018multi} showed that this process is highly beneficial for mitigating the errors due to variation of view angles in gait recognition through view synthesis.

For benefit of the readers who may be unfamiliar with GAN we provide a brief introduction. Generative adversarial networks fall within the family of generative models. The Generator ($G$) learns a mapping from a random noise vector $z$ to an output $y, G: z \rightarrow y$ \cite{goodfellow2014generative}. An extension to this basic model is proposed in \cite{isola2017image} where the authors propose a conditional GAN, which learns a mapping from an observed input $x$ and random noise vector $z$, to output $y$, $G: \{z, x \} \rightarrow y$. This extension allows the model to learn a conditional mapping between the current input and the output. 

GANs partake in a two-player adversarial game where the generator, $G$, tries to fool the discriminator, $D$, with synthesised outputs while $D$ tries to identify them. This objective, in terms of the conditional GAN, can be written as,  
\begin{equation}
\mathop{min}_{G} \mathop{max}_{D} \hspace{1mm}\mathbb{E}_{x, y}[log (D(x, y))] + \mathbb{E}_{x, z}[log(1- (D(x, G(x, z))))],
\end{equation}
where $D$ tries to maximise this objective while $G$ tries to minimise it. Hence there exists a dual between $G$ and $D$, through which the GAN framework learns a custom loss function for the task at hand. It should be noted that we do not explicitly define the loss of $G$. The discriminator, $D$, is the loss function for the $G$, which is a neural network approximating the loss. Therefore, a custom loss function is learned through the adversarial learning process. For further information regarding the GAN learning process we refer the readers to \cite{goodfellow2014generative,isola2017image}.

GANs are extensively applied for tasks such as image-to-image synthesis \cite{isola2017image,yi2017dualgan,liu2017unsupervised,wang2017high}, video synthesis \cite{berthelot2017began,wang2018video} and speech enhancement \cite{donahue2018exploring,pascual2017segan,soni2018time}, but seldom for SAD. To the best of our knowledge, no prior work has applied GANs for the SAD task. 
Most GAN related works have focused on using static inputs such as images \cite{isola2017image,yi2017dualgan,liu2017unsupervised,wang2017high}, while only a few have addressed temporal changes in the input data. In \cite{saito2017temporal,yu2017seqgan} the authors address this by directly incorporating the time axis in the input and output. For instance, in \cite{saito2017temporal} the authors propose a temporal generator while Yu et. al \cite{yu2017seqgan} propose a sequence generator that learns a stochastic policy. However, neither of these works have considered a framework that processes individual frames while also considering the temporal relationships between them. 

Xie et. al \cite{xie2018tempogan} address this issue through a dual discriminator architecture. However, they have engineered the temporal loss to consider the velocity of consecutive frames, and hence this cannot be directly applied for speech processing. 

In our work we exploit the merits of the GAN learning framework  to automatically learn a loss function for synthesising highly indistinguishable data and synthesise both the speech activity classifications for the set of individual input frames as well as the input signal in the next time frame. This allows us to learn the context of the input audio segment. 

In the context of computer vision, $L_1$ and $L_2$ losses have been extensively coupled with the adversarial GAN loss to alleviate the static pixel-wise loss between the synthesised output and the ground truth data \cite{dong2016image,isola2017image}. The $L_2$  loss minimises the sum of the squared differences between the synthesised output and ground truth data \cite{steinier1972smoothing} while the $L_1$ loss function minimises the sum of the absolute differences between the synthesised output and ground truth data \cite{bloomfield1984least}. In \cite{berthelot2017began} the authors demonstrate that $L_2$ loss is more effective in penalising discontinuities between nearby frames compared to the $L_1$ loss. Motivated by these findings we utilise $L_2$ as our regularisation mechanism.

\section{Architecture}

The proposed architecture is inspired by the success of multi-task learning over single task learning methods in numerous speech related areas \cite{pironkov2016multi, kim2017speech, chen2015speech}. We design our auxiliary task of predicting the next audio segment to facilitate our primary goal of speech/non-speech classification via capturing broader context of the input audio segment than just relying on the input itself. Rather than using hand engineered loss function for the two tasks we utilise GAN framework to automatically learn a custom loss function that facilitates both tasks.

The proposed approach is shown in Fig. \ref{fig:TA_GAN_model}. Inputs are processed by the encoder, $f^E$, which embeds this information into a vector. We implemented the encoding function $f^E$ using a single LSTM cell. Using this embedding, the generator, $G^{\eta}$, synthesises a speech activity classification sequence while $G^{w}$ synthesises the future audio signal (see Sec. \ref{sec:generators}). We utilise two discriminators, a static discriminator (see Sec. \ref{sec:st_disc}) and a temporal discriminator (see Sec. \ref{sec:temp_disc}) where the former considers individual elements in the sequence when performing the adversarial classification, and the latter preserves the temporal relationships between audio frames of the outputs. The overall objective of the combined model is presented in Sec. \ref{sec:overall}. 

Motivated by \cite{mccowan2011delta} we consider a combination of input features. Let the input, $X$, be,
\begin{equation}
X= [(w_1,w_2, \ldots, w_T), (\theta_1,\theta_2, \ldots, \theta_T), (\Delta_1, \Delta_2, \ldots, \Delta_T)], 
\end{equation}

where $w_t$ is the raw audio of the frame $t$, $\theta_t$ denotes the MFCC feature \cite{mccowan2011delta} and $\Delta_t$ denotes the MFCC deltas \cite{zhang2018speech} for the same frame. 

\subsection{Generators}
\label{sec:generators}
Given an input $X$, we first pass it through an encoding function, $f^E$, which generates an embedding such that,
\begin{equation}
C=f^{E}(X),
\end{equation}
where $C=[c_1, c_2, \ldots, c_t, \ldots, c_T]$. Using this input embedding, $C$, and a noise vector, $z$, the generator, $G^{\eta}$, synthesises a speech classification sequence, $\hat{\eta} =[ \hat{\eta}_{1},\hat{\eta}_{2}, \ldots, \hat{\eta}_{t}, \ldots, \hat{\eta}_{T}] $, classifying each frame in $X$, while $G^{w}$ synthesises the future audio signal, $\hat{w}=[\hat{w}_{T+1}, \hat{w}_{T+2}, \ldots, \hat{w}_{T+T}]$, for the next time window. This can be written as,
 \begin{equation}
 \hat{\eta}= G^{\eta}(C,  z),
\end{equation}
and
 \begin{equation}
\hat{w}= G^{w}(C, z).
\end{equation}
Predicting the raw signal, rather than MFCC's or other features, allows us to enforce temporal constraints in the discriminator and preserve the original characteristics of the input signal. 

\subsection{Static Discriminator}
\label{sec:st_disc}
The static discriminator, $D^{\eta}$, receives the current input embeddings and the ground truth speech classification sequence, $\eta$, and learns to classify it as real while $G^{\eta}$ tries to synthesise a classification sequence, $\hat{\eta}$, which is not easily distinguishable from the real sequences. This objective can be written as, 
\small
\begin{equation}
\begin{split}
V^{\eta}= \mathop{min}_{G^{\eta}} \mathop{max}_{D^{\eta}}  \sum_{t=1}^{T}\hspace{1mm}\mathbb{E}[log (D^{\eta}(c_t, \eta_t))] \\
+ \sum_{t=1}^{T}\mathbb{E}[log(1- (D^{\eta}(c_t, G^{\eta}(c_t, z))))] + \lambda^{\eta} \sum_{t=1}^{T}||\eta_t - G^{\eta}(c_t, z) ||^{2} ,
\end{split}
\label{eq:static_dis}
\end{equation}
\begingroup
\fontsize{10pt}{12pt}\selectfont
where we add an additional $L_2$ loss to regularise the process and $\lambda^{\eta}$ is a hyper-parameter controlling the contribution from the $L_2$ loss. 
\endgroup

\subsection{Temporal Discriminator}
\label{sec:temp_disc}
\begingroup
\fontsize{10pt}{12pt}\selectfont
The objective in Eq. \ref{eq:static_dis} is shown to be highly effective for generating realistic static outputs considering the elements of the sequence individually \cite{bansal2018recycle}. However, it discards the temporal coherence as the generator and the discriminator consider each frame individually \cite{bansal2018recycle}. 
Even though this behaviour is acceptable when considering the frame-wise speech classification sequence, it is suboptimal when considering the future audio output. Inspired by \cite{bansal2018recycle,xie2018tempogan,yu2017seqgan} we introduce a temporal discriminator, $D^{w}$, which also preserves the temporal relationships between audio frames of the output. 
We consider different sub sequences of the generated sequences $\hat{\eta}$ and $\hat{w}$, and generate the true/ fake classification through the discriminator considering these sub-sequences. Hence it forces the discriminator to consider the temporal accordance of these sub-sequences. This objective can be written as, 
 \begin{equation}
\begin{split}
 V^{w}=\mathop{min}_{G^{w}} \mathop{max}_{D^{w}}  \sum_{t=1}^{T}\mathbb{E}[log (D^{w}(c_{1 : t}, w_{t+1 : t+t}))] \\
 + \sum_{t=1}^{T} \mathbb{E}[log(1- (D^{w}(c_{1 : t}, G^{w}(c_{1 : t},z)))) ]\\
 + \lambda^{w}\sum_{t=1}^{T} ||w_{t+1 : t+t} - G^{w}(c_{1 : t},z) ||^{2},
 \end{split}
  \label{eq:tempo_dis}
\end{equation}
where $w_{1:t}= (w_1 \ldots w_t)$, $c_{1 : t}= f^{E}(X_{1:t})$ and $\lambda^{w}$ is a hyper-parameter controlling the contribution from the $L_2$ loss. 
\endgroup

We would like to emphasise the fact that utilising the above formulation, the static discriminator provides frame-wise true/fake decisions while the temporal discriminator provides decisions for time-windows of different frame lengths.

\subsection{Complete Model}
\label{sec:overall}
We combine the objectives in Equations \ref{eq:static_dis} and \ref{eq:tempo_dis} to obtain the objective for the proposed TA-GAN,
 \begin{equation}
 V^*=V^{\eta} + V^{w}.
\end{equation}
It can be seen that for the individual losses $V^{\eta}$ and $V^{w}$ there exist contributions from the adversarial losses which occur due to the dual between  $G^{\eta}$ and $D^{\eta}$, and $G^{w}$ and $D^{\eta}$. As shown in Equations \ref{eq:static_dis} and \ref{eq:tempo_dis}, the generators $G^{\eta}$ and $G^{w}$ try to minimise these loss values while discriminators $D^{\eta}$ and $D^{\eta}$ try to maximise them. Hence it can be concluded that the overall loss, $V^*$, of the proposed TA-GAN is automatically learned through the proposed framework by considering the task at hand. 

\section{Evaluations}

 \subsection{Datasets}
The proposed Temporarily-Aware GAN (TA-GAN) framework is evaluated on four popular SAD benchmarks, namely, HAVIC \cite{havic_ldc}, AMI Meeting corpus \cite{carletta2006announcing}, NIST OpenKWS'13 \cite{harper2014iarpa}, and NIST OpenSAT' 17 \cite{nist_open_SAT}. The details of the datasets and the evaluation protocols are summarised below.

\subsubsection{HAVIC}
HAVIC  (the Heterogeneous Audio Visual Internet Collection) Pilot Transcription \cite{havic_ldc} is comprised of approximately 72 hours of user-generated videos with transcripts based on the English speech audio extracted from the videos. The transcription files contain the type of the audio segment annotated for speech, music, noise and singing segments \cite{himawan2018investigating}. We choose music and noise segments as non-speech and rest of the segments as speech. Due to the unavailability of standard training/ testing splits we randomly split 70\% of the data for training, 20\% for testing and 10\% for validation.  As the evaluation metric we measure NIST OpenSAD Detection Cost Function (DCF),

\begin{equation}
DCF=0.75 \times P_{miss} + 0.25 \times P_{fa},
\label{eq:dcf}
\end{equation}
where $P_{miss}$ denotes miss probability and $ P_{fa}$ denotes the probability of false alarms. 

\subsubsection{AMI Meeting corpus}
This dataset consists of 100 hours of recordings collected across three different meeting rooms. It offers a challenging SAD setting as audio data is from both non-native and native English speakers. Similar to \cite{gelly2018optimization} we use the Frame Error Rate (FER) metric to evaluate the performance. Training testing splits are as defined in \cite{carletta2006announcing}.

\subsubsection{NIST OpenSAT' 17}
We also utilise the public safety communications (PSC) corpus from NIST OpenSAT 2017 for our evaluations \cite{nist_open_SAT},  which is a standard split in NIST OpenSAT 2017 and is constructed using the audio data from Sofa Super Store Fire (SSSF) dispatcher that occurred on June 18, 2007 in Charleston, South Carolina. This data consisted of audio logs in English from real fire-response operational data and is rich in naturalistic distortions including land-mobile-radio transmission effects, speech under cognitive and physical stress, speaking with significant background noise (Lombard effect), varying background-noise types and levels, and varying background decibel levels, \cite{nist_open_SAT, dubey2018leveraging,byers20192017}. The data is provided as 16-bit at 8 kHz sampling rate \footnote{We obtained the data from https://catalog.ldc.upenn.edu/ and the LDC Catalog ID is LDC2017E12}. Due to the unavailability of ground truth evaluation labels we use the six audio recordings in the development data which constitute approximately 30 minutes worth audio recordings. Due to this limited size we utilise this dataset only under cross database evaluations (see Sec. \ref{sec:cross_database}) where we use this dataset only for testing (i.e it is not used for training the models). Following \cite{dubey2018leveraging} we measure the DCF metric which is evaluated using Eq. \ref{eq:dcf}. 

\subsubsection{NIST OpenKWS'13}
To demonstrate the robustness of TA-GAN for different languages we evaluate the performance using Vietnamese, Pashto, Turkish and Tagalog corpuses from the IARPA Babel dataset \cite{harper2014iarpa} \footnote{Vietnamese IARPA-babel107b-v0.7, Pashto IARPA-babel104b-v0.4b,Turkish IARPA-babel105b-v0.5, Tagalog IARPA-babel106-v0.2g}. We evaluate the system using the FER metric as in \cite{gelly2018optimization}.

\subsection{Implementation Details}
We use a sliding window \cite{pascual2017segan} to sample 1 second segments from the raw audio every 500ms (with 50\% overlap). We extract MFCC features with 13 cepstral coefficients and the delta features considering the immediately preceding 2 frames and the next 2 frames using a frame size of 25 ms, sampled at a frame rate of 100 fps. Similar to \cite{nagrani2017voxceleb} inputs are normalised to a range 0-1, and no other speech-specific preprocessing is performed. At test time we slide the window, without overlap, over the whole test utterance and generate the relevant speech classification sequence using $G^{\eta}$. It should be noted that similar to \cite{gelly2018optimization} we generate speech/ non- speech predictions for each frame within the 1 second segment. Hence at test time there is only a 1 second framing delay at the beginning, after which the window can be shifted in small increments to produce predictions in real time. 
 
\begin{figure}[htbp]
\centering
\subfigure[$\lambda^{\eta}$ vs FER]{\includegraphics[width = .45 \textwidth]{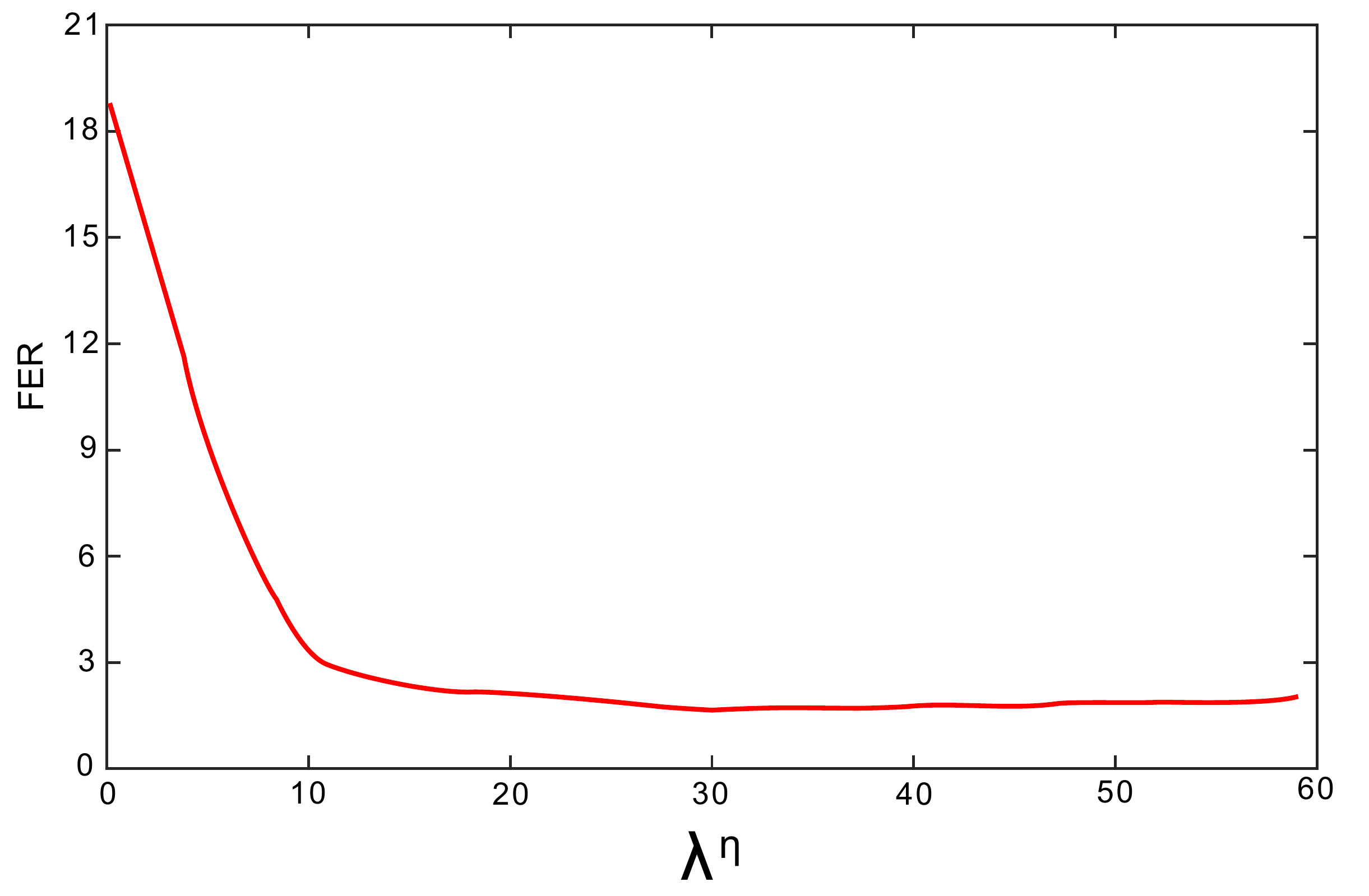}}
\subfigure[ $\lambda^{w}$ vs FER]{\includegraphics[width = .45 \textwidth]{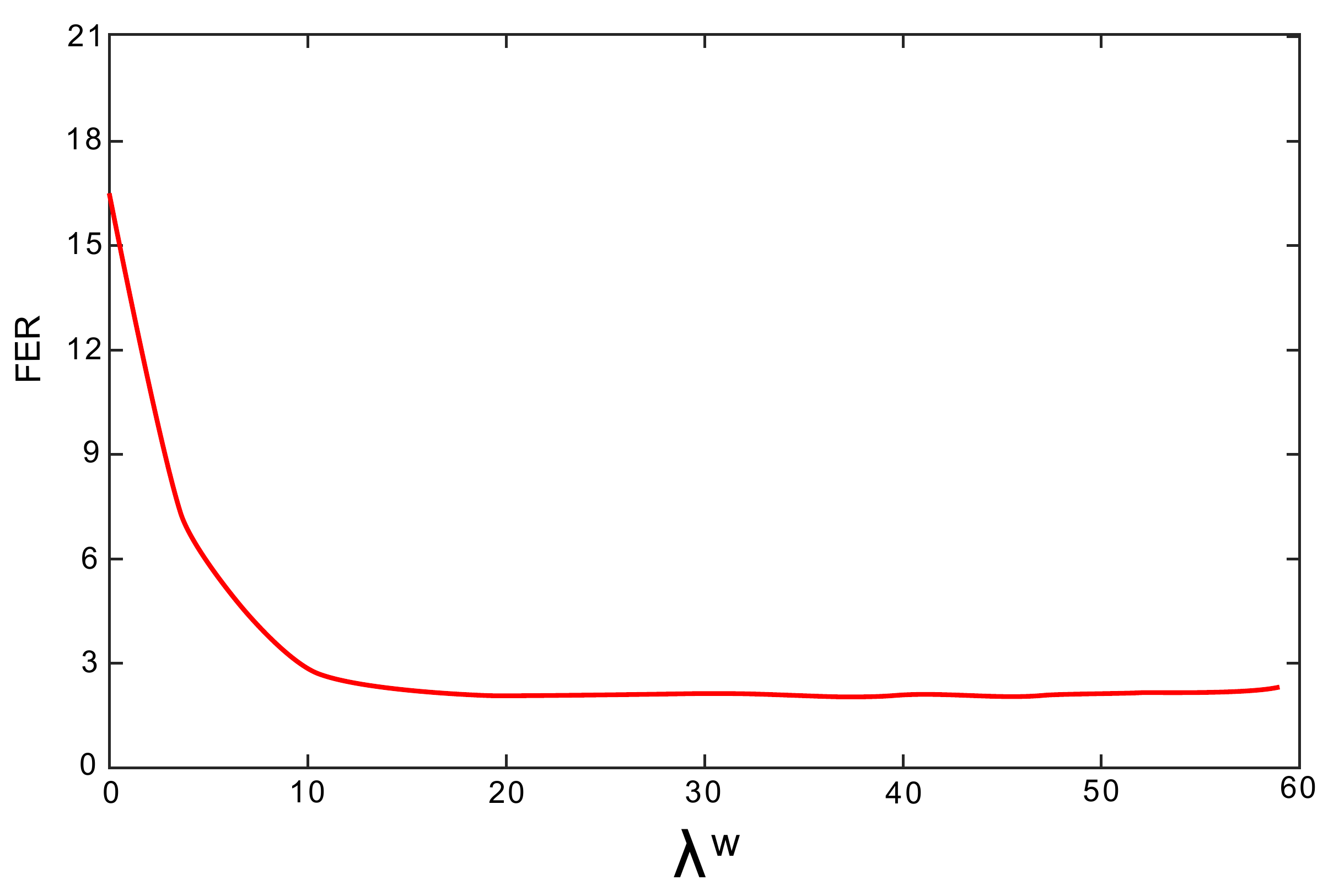}}
\caption{Evaluation of hyper-parameters using the validation set of AMI Meeting corpus. We set $\lambda^{\eta}=30$ and $\lambda^{w}=25$. }
\label{fig:hyper_parameters}
\end{figure}

We implemented the encoding function, $f^E$, using a single LSTM cell, and the two generators, $G^{\eta}$ and  $G^{w}$, are implemented with two separate LSTM blocks, each with two cells of LSTMs. For all LSTMs the hidden state size is set to 300 units. For training, we use the Adam \cite{kinga2015method} optimiser, a learning rate of 0.005, and 500 epochs with a batch size of 600, alternating between epochs of $D$ and $G$. We train the input encoder jointly with the two generators $G^{\eta}$ and $G^{w}$. However the two discriminators $D^{\eta}$ and $D^{w}$ are updated individually.  Hyperparameters $\lambda^{\eta}$ and $\lambda^{w}$ are evaluated experimentally by changing the respective hyper-parameter while holding the rest of the parameters constant, and are set to 30 and 25 respectively. Changes in FER against $\lambda^{\eta}$ and $\lambda^{w}$ are shown in Fig. \ref{fig:hyper_parameters}. The implementation of the proposed TA-GAN is completed with Keras \cite{chollet2017keras} and Theano \cite{bergstra2010theano}.

\subsection{Results}

\begin{table}[htbp]
\caption{Evaluations on the HAVIC dataset \cite{havic_ldc}. DCF denotes NIST OpenSAD Detection Cost Function (DCF) as defined in Eq. \ref{eq:dcf}.}
\centering
\resizebox{.65\linewidth}{!}{
\begin{tabular}{|c|c|}
\hline
Method              & DCF \\\hline
MLP - Gelley et. al \cite{gelly2018optimization} & 8.10 \\
Basic RNN - Gelley et. al \cite{gelly2018optimization}    & 6.38 \\
CG-LSTM - Gelley et. al \cite{gelly2018optimization}& 5.10    \\ 
ACAM -Kim et al - \cite{kim2018voice} & 4.95 \\  \hline
TA-GAN              & \textbf{2.53} \\
\hline
\end{tabular}
}
\label{tab:havic}
\end{table}

\begin{table}[htbp]
\caption{Evaluations on the AMI Meeting \cite{carletta2006announcing}  and OpenKWS'13 corpus \cite{harper2014iarpa}. FER denotes Frame Error Rate as defined in \cite{gelly2018optimization}.}
\centering
\resizebox{1\linewidth}{!}{
\begin{tabular}{|c|c|c|}
\hline
\multirow{2}{*}{Method} & \multicolumn{2}{c|}{FER}  \\ \cline{2-3} 
                        & AMI Meeting & OpenKWS'13 \\ \hline
MLP - Gelley et. al \cite{gelly2018optimization}    & 6.84        & 6.29        \\ 
Basic RNN - Gelley et. al \cite{gelly2018optimization}    & 6.55 & 6.24 \\
CG-LSTM - Gelley et. al \cite{gelly2018optimization}& 5.93        & 5.76        \\ 
ACAM - Kim. et al \cite{kim2018voice} & 5.89 & 5.66 \\  \hline
TA- GAN                 & \textbf{2.80}        &     \textbf{2.75}         \\ \hline
\end{tabular}
}
\label{tab:ami_and_bable}
\end{table}

Evaluations on the HAVIC dataset are presented in Tab. \ref{tab:havic}, and AMI Meeting corpus and NIST OpenKWS'13 corpora are presented in Tab. \ref{tab:ami_and_bable}. For better comparisons we provide evaluations for the CG-LSTM and Basic RNN and MLP methods in \cite{gelly2018optimization}, and the Adaptive Context Attention Model (ACAM) proposed in \cite{kim2018voice}. These two models which were proposed very recently have been able to attain state-of-the-art results under a supervised SAD setting in the datasets that we consider.  The work of Gelley et. al \cite{gelly2018optimization} utilises RNNs for modelling the temporal relationships within the input signal and demonstrates that directly optimising the SAD loss using the Frame Error Rate (FER) produces better results. In \cite{kim2018voice} Kim. et al exploit an attention strategy for learning the context of the speech signal using LSTMs for noise robustness in the SAD system. The comparative evaluations with these baselines demonstrates the utility of GAN based learning for the SAD system. In addition to utilising LSTMs for temporal modelling and an attention mechanism for input embedding, the proposed model automatically learns a loss function for the SAD task. Hence, in contrast to \cite{gelly2018optimization, kim2018voice}, the proposed TA-GAN method has been able to learn a more robust input embedding which better discriminates the speech segments compared to its counterparts. Furthermore, when comparing the MLP - Gelley et. al \cite{gelly2018optimization} system with Basic RNN - Gelley et. al \cite{gelly2018optimization}, CG-LSTM - Gelley et. al \cite{gelly2018optimization}, recurrent neural network based temporal modelling has been able to further improve the performance over an MLP network. We would like to note that these systems directly optimise the FER loss. In contrast, using the task-specific loss function learning framework of GANs and the augmented multi-task learning approach, the proposed method has been able to outperform the state-of-the-art methods. 

\subsection{Cross Database Evaluation}
\label{sec:cross_database}
To demonstrate the robustness of the proposed method across different languages, accents, and acoustics, we perform a cross-database evaluation where we train the model using the training data of one dataset and test that model on the test sets of the rest of the datasets. 

The evaluations are presented in Tab. \ref{tab:cross_database}. Note that when tested on NIST OpenSAT' 17 \cite{nist_open_SAT} and HAVIC datasets \cite{havic_ldc} we report the NIST OpenSAD Detection Cost Function (DCF) whereas for AMI Meeting and OpenKWS'13 corpus we report the Frame Error Rate (FER). To better demonstrate the merits of the proposed method we train the CG-LSTM baseline model defined in \cite{gelly2018optimization} and ACAM baseline model of \cite{kim2018voice}. For better comparisons for AMI Meeting, HAVIC and OpenKWS'13 datasets, within brackets we report the error rates when the model is trained and tested on the database indicated in the ``Tested On" column. Due to the limited dataset size we do not attempt to train models using the NIST OpenSAT' 17 dataset. 

When analysing the results it is clear that the proposed GAN based learning framework better captures the discriminative features and is more robust under cross domain scenarios, better segregating speech from non-speech embeddings. This allows the proposed method to achieve superior results compared to the baselines. Comparing the cross database evaluations with the evaluations presented within brackets, when the models are trained and tested on the same dataset, we only observe a slight reduction in the performance of the proposed approach when it is not tuned on the training set of the specific dataset. However, the performance reductions in the baselines are quite substantial.

\begin{table}[htbp]
\caption{Cross database evaluations using NIST OpenSAT' 17 \cite{nist_open_SAT}, HAVIC \cite{havic_ldc}, AMI Meeting \cite{carletta2006announcing} and OpenKWS'13 corpus \cite{harper2014iarpa}. For  HAVIC, AMI Meeting and OpenKWS'13 datasets within brackets we report the error rates when the model is trained and tested on the database indicated in the ``Tested On" column.}
\centering
\resizebox{\linewidth}{!}{
\begin{tabular}{| c | c | c | c | c |}
\hline
                          &                            & \multicolumn{3}{c|}{ Error Rate (DCF / FER)} \\ \cline{3-5} 
\multirow{-2}{*}{Trained on}  & \multirow{-2}{*}{Tested on} & CG-LSTM \cite{gelly2018optimization}              & ACAM \cite{kim2018voice}             & TA- GAN                    \\ \hline
                    & NIST OpenSAT' 17                                    & 5.36               & 4.78              & \textbf{2.53}              \\ \cline{2-5} 
                        & HAVIC                                               & 7.63 (5.10)        & 7.08 (4.95)       & \textbf{4.53 (2.53)}       \\ \cline{2-5} 
\multirow{-3}{*}{AMI Meeting} & OpenKWS'13                                          & 8.17 (5.76)        & 7.73 (5.66)       & \textbf{4.15 (2.75)}       \\ \hline \hline
                            & NIST OpenSAT' 17                                    & 5.30               & 4.42              & \textbf{2.14}              \\ \cline{2-5} 
                              & OpenKWS'13                                          & 7.93 (5.76)        & 7.65 (5.66)       & \textbf{4.01 (2.75)}       \\ \cline{2-5} 
\multirow{-3}{*}{HAVIC}       & AMI Meeting                                         & 7.87 (5.93)        & 7.39 (5.89)       & \textbf{4.23 (2.80)}       \\ \hline \hline
                             & NIST OpenSAT' 17                                    & 5.51               & 5.02              & \textbf{3.14}              \\ \cline{2-5} 
                      & HAVIC                                               & 7.86 (5.10)        & 7.12 (4.95)       & \textbf{4.81 (2.53)}       \\ \cline{2-5} 
\multirow{-3}{*}{OpenKWS'13}  & AMI Meeting                                         & 8.51 (5.93)        & 7.71 (5.89)       & \textbf{4.60 (2.80)}       \\ \hline 
\end{tabular}}
\label{tab:cross_database}
\end{table}

We note that the baseline ACAM \cite{kim2018voice} model utilises a window of 39 frames as the input, $w$, while the proposed TA-GAN model utilises 100 frames as the input window. Due to this difference, their performance is not directly comparable. However, in Sec. \ref{sec:window_size} we show a further evaluation using different (smaller) window sizes which illustrates that the proposed TA- GAN model is capable of outperforming the baseline models even with smaller input window sizes.

\subsection{Ablation Experiments}
\label{sec:ablation}
To better understand the crucial components and sensitivities of the proposed TA-GAN framework, we conduct a series of ablation experiments. In this experiment, we use the AMI Meeting \cite{carletta2006announcing} dataset and compare the TA-GAN model with a series of counterparts defined as follows:

\begin{enumerate}
  \item $G^{\eta}(w)$: Removes the GAN learning framework and $G^{\eta}$ is learnt through binary cross entropy loss. This receives only the audio input.
    \item $G^{\eta}(w+ \theta + \Delta)$: Receives audio, MFCC and delta inputs.
    \item $G^{\eta} + G^{w} (w+ \theta + \Delta)$: Similar to 2) but additionally predicts the future audio segment, which is trained using mean square error.   
    
    \item $GAN^{\eta}(w+ \theta + \Delta) / L_2$: uses the GAN learning framework but synthesises only the classification sequence. Receives audio, MFCC and delta inputs. Doesn't utilise $L_2$ regularisation in Eq. \ref{eq:static_dis}
    \item $GAN^{\eta}(w+ \theta + \Delta)$: Same as above method but with $L_2$ regularisation.
    
   \item $TA-GAN(w)$: Proposed model that receives only the audio input and predicts the future audio segment.
   \item $TA-GAN(\theta)$: Proposed model which receives the MFCC as input and predicts future MFCC distribution.
  \item $TA-GAN(\Delta)$: Receives Deltas of MFCC as input and predicts future deltas.
  \item $TA-GAN(w + \theta)$: Receives both audio and MFCC inputs and predicts their future distributions.
  \item $TA-GAN(w + \Delta)$: Receives both audio and delta inputs and predicts their future distributions.
    \item $TA-GAN(\theta + \Delta)$: Receives both MFCC and delta inputs and predicts their future distributions.
    \item $TA-GAN(w+ \theta + \Delta) / L_2$: Receives audio, MFCC and delta inputs and predicts their future distributions. Doesn't utilise $L_2$ regularisation in Eq. \ref{eq:static_dis} and Eq. \ref{eq:tempo_dis}.
     \item $TA-GAN (\dot{D}^{w})$: Replaced the temporal discriminator with static discriminator as per Eq. \ref{eq:static_dis}, hence, this model contains two static discriminators.
    
\end{enumerate}

\begin{table}[htbp]
\caption{Ablation model evaluations on AMI Meeting \cite{carletta2006announcing} dataset.}
\centering
\resizebox{.75\linewidth}{!}{
\begin{tabular}{|c|c|c|}
\hline
ID & Method              & FER \\\hline
1) &$G^{\eta}(w)$         & 9.10 \\
2)& $G^{\eta}(w+ \theta + \Delta)$      & 7.20  \\
3)& $(G^{\eta} + G^{w}) (w+ \theta + \Delta)$      & 7.12  \\\hline

4)& $GAN^{\eta}(w+ \theta + \Delta) / L_2$ & 4.73 \\
5)& $GAN^{\eta}(w+ \theta + \Delta)$      &   4.15\\\hline

6)& $TA-GAN(w)$         &  3.99\\
7)  &$TA-GAN(\theta)$       &  3.72\\
8) &$TA-GAN(\Delta)$    &  4.03\\
9)& $TA-GAN(w + \theta)$       &  3.60\\
10) &$TA-GAN(w + \Delta)$ &   3.98\\
11) &$TA-GAN(\theta + \Delta)$ & 3.65\\
12) &$TA-GAN(w+ \theta + \Delta) / L_2$ & 3.54\\
13) &$TA-GAN (\dot{D}^{w})$    &   3.31 \\ \hline

\textbf{Proposed} &\textbf{TA-GAN}              & \textbf{2.80} \\
\hline
\end{tabular}
}
\label{tab:ablation}
\end{table}

With the ablation evaluations presented in Tab. \ref{tab:ablation} we can see the importance of multi-task learning, the merits of using GAN based automatic loss function learning and the importance of utilised features. 

When comparing both non-GAN and GAN based single task SAD methods (ablation model 1-2 and 4-5) with their respective multi-task counterparts (i.e ablation models 3 and 13) we observe a significant contribution for the SAD task through the multi-task learning strategy.  Furthermore, when comparing non-GAN based models (1-3) with GAN based models (4-13), we observe a significant performance boost denoting the merits of task-specific loss function learning. We would like to emphasise the fact that this performance increase is observed for both single-task as well as multi-task models, although we observe a further substantial improvement with regards to multi-task methods.

In addition we observe that the temporal discriminator has been able to further improve this learning process (see model 13 and TA-GAN (proposed)). Even though we do not observe a direct relationship between the temporal discriminator, which is used for real/fake validation of the predicted future audio segments and the SAD task, we notice a significant contribution from this module. This illustrates that via analysing the temporal relationships between audio frames the discriminator gains the ability to guide the generator to generate realistic outputs. Hence it enforces the input embeddings to better identify the temporal context of the inputs, denoting the utility of multi-task learning and the importance of the future audio segment prediction task as the auxiliary task of the proposed multi-task learning framework. 

When comparing different feature combinations present in Tab. \ref{tab:ablation} we observe that MFCC features contain more salient attributes for the SAD task. However, when MFCC features are fused with both audio and $\Delta$ features we observe improved performance, highlighting that the complementary attributes present in those streams have the ability to better discriminate speech segments from their counterparts. 

\subsection{Impact of input window size}
\label{sec:window_size}

In order to illustrate the impact of the input window size for SAD accuracy , we perform an additional evaluation on the proposed TA-GAN model using different window sizes: 20, 40, 60, 80, 100, and 120 frames. In this experiment, we use the AMI Meeting \cite{carletta2006announcing} dataset.

\begin{table}[htbp]
\caption{Evaluating the effect of different window sizes on the TA-GAN model using AMI Meeting \cite{carletta2006announcing} dataset.}
\centering
\begin{tabular}{|c|c|}
\hline
Window Size (in frames) & FER  \\ \hline
20                      & 5.24 \\ \hline
40                      & 4.41 \\ \hline
60                      & 3.54 \\ \hline
80                      & 3.11 \\ \hline
100                     & 2.80 \\ \hline
120                     & 2.84 \\ \hline
\end{tabular}
\end{table}

Considering these evaluations it is clear that a considerable reduction in the FER can be achieved when increasing the window size from 20 to 100 frames, but no significant gain is observed by increasing it beyond 100 frames. We believe utilising a large window size is essential in the proposed method in order to properly model the context within the given window. Furthermore, comparing these results to those obtained by ACAM \cite{kim2018voice} for the AMI Meeting using a window size of 39 frames, we observe that the proposed TA-GAN model with a smaller window size (i.e 20 frames) has been able to achieve better performance than ACAM \cite{kim2018voice}.

\subsection{Qualitative Results}
\begin{figure}[htbp]
\centering
\subfigure[TA-GAN]{\includegraphics[width = .6 \linewidth]{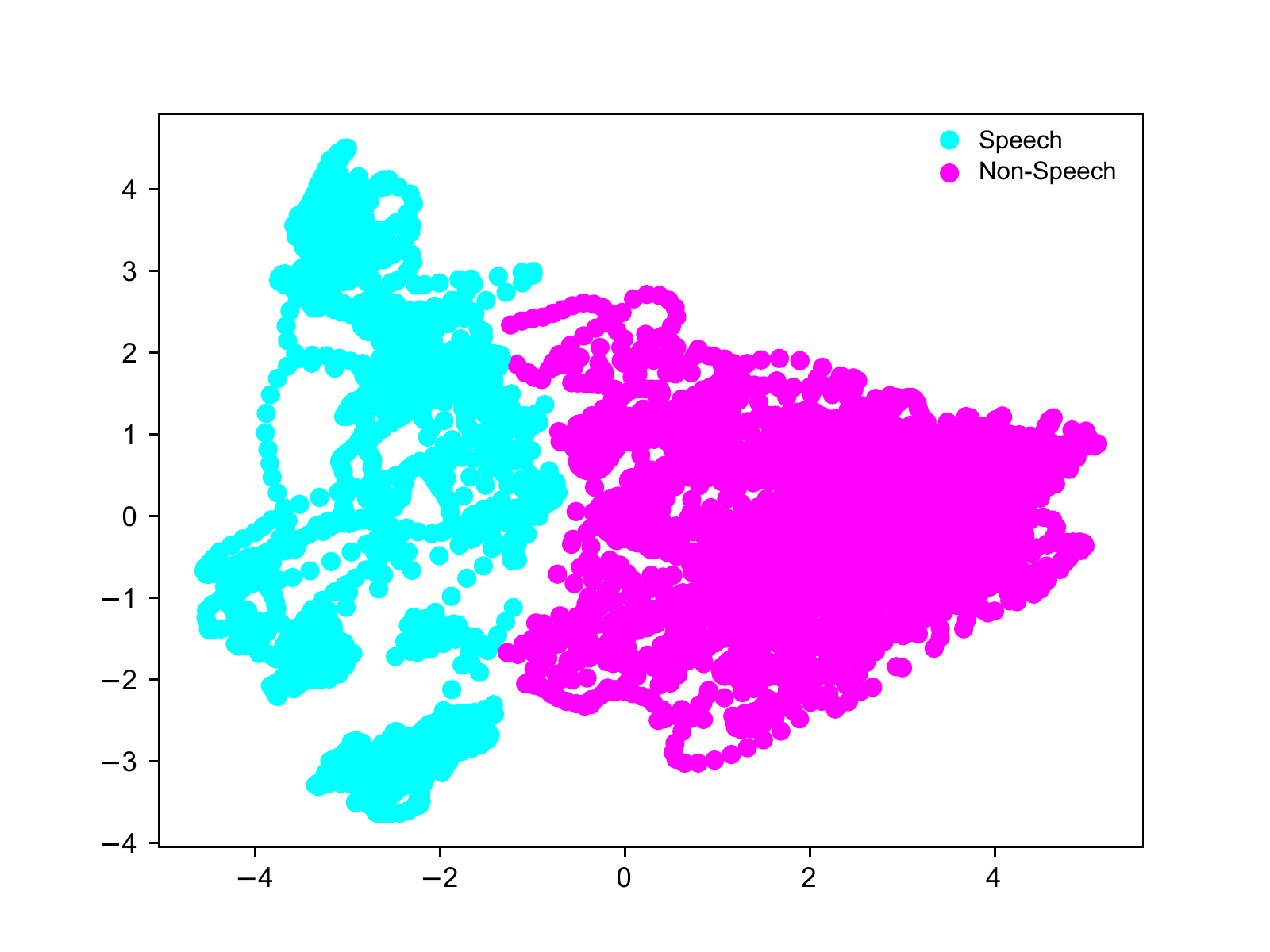}}
\subfigure[ Ablation model 3 ($G^{\eta} + G^{w} (w+ \theta + \Delta)$) ]{\includegraphics[width = .6 \linewidth]{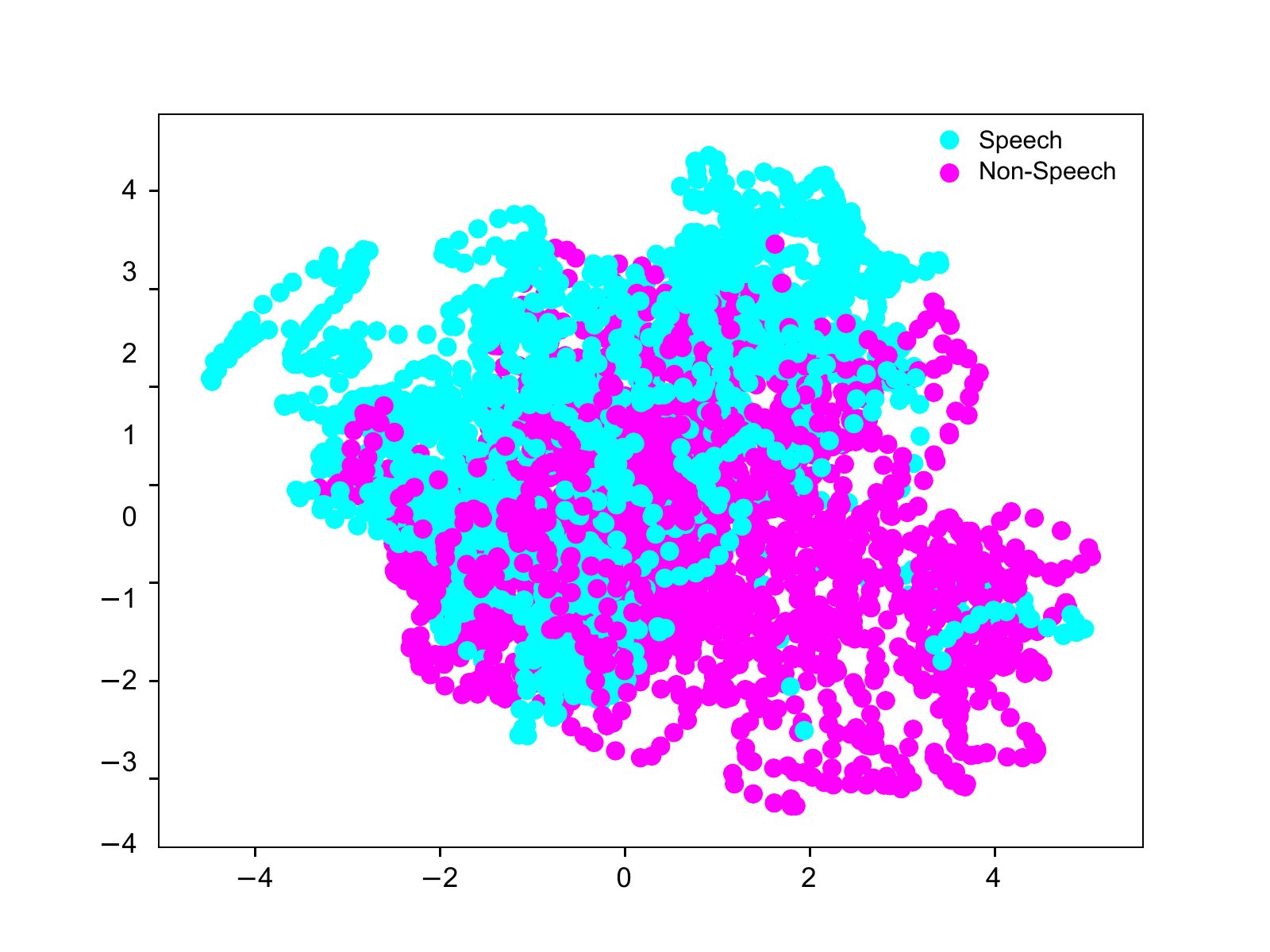}}
\subfigure[ Ablation model 5 ($GAN^{\eta}(w+ \theta + \Delta)$) ]{\includegraphics[width = .605 \linewidth]{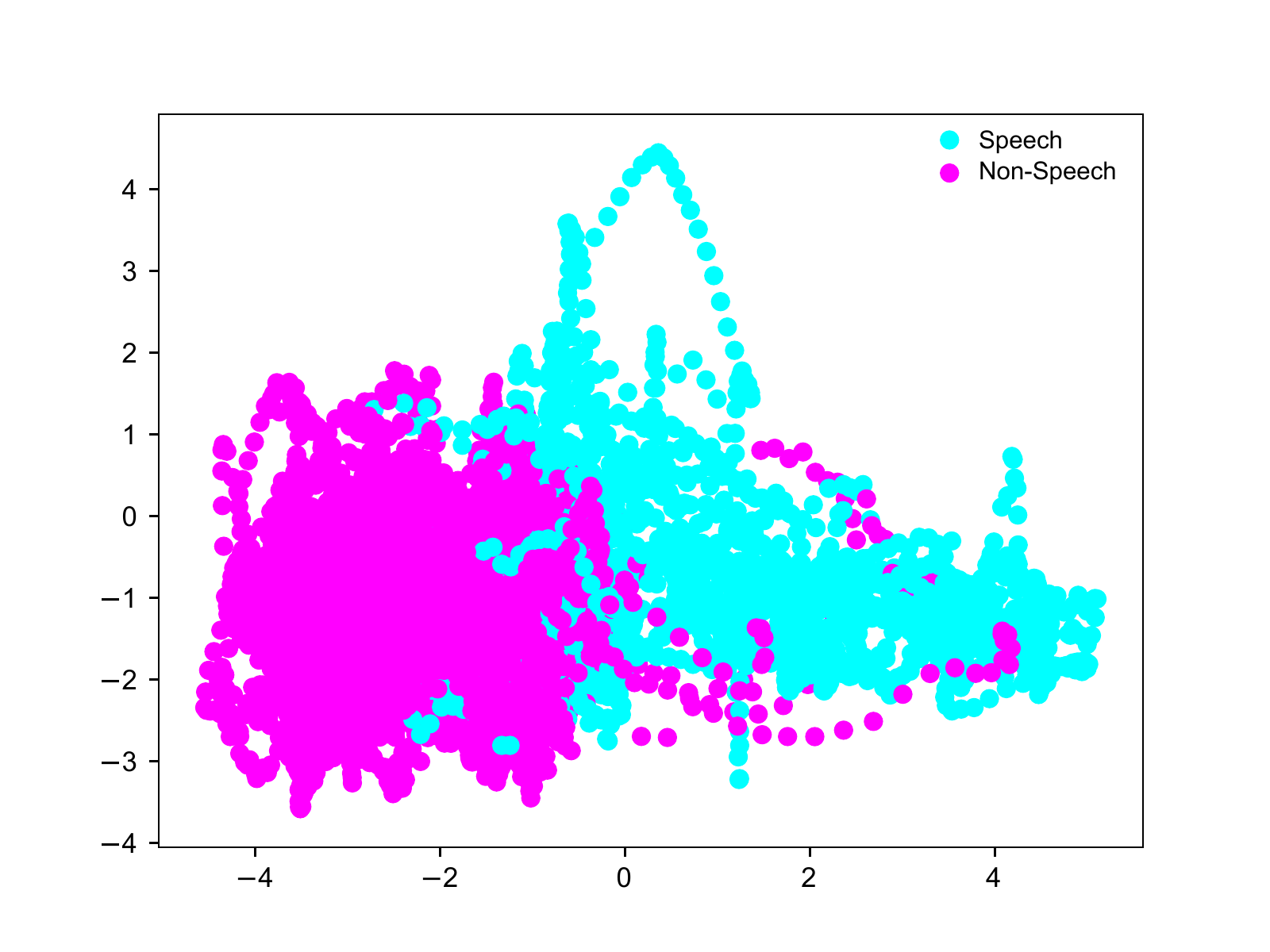}}
\caption{Visualisation of input embeddings, $C$,  for proposed TA-GAN and Ablation models  3 ($G^{\eta} + G^{w} (w+ \theta + \Delta)$) and 5 ($GAN^{\eta}(w+ \theta + \Delta)$)}
\label{fig:context_embeddings}
\end{figure}

We randomly selected 100 examples from the AMI Meeting \cite{carletta2006announcing} test set and plotted the inputs embeddings, $\mathbf{c}$, for each frame in those examples. The model trained on the AMI Meeting training set is used generate these embeddings. These embeddings are coloured based on the ground truth speech/ non-speech labels. Note that for each frame the encoder generates a 300 dimensional embedding vector. Hence in order to plot the results in 2D we applied PCA \cite{wold1987principal} to reduce the dimensionality. In Fig. \ref{fig:context_embeddings} (a) we visualise the input embeddings learnt through the proposed TA-GAN model. It is clear that the model has been successful in learning an embedding space which better segregates speech from non-speech than the alternate ablation models, at least in terms of the two directions that capture most variation determined via PCA. In Fig, \ref{fig:context_embeddings} (b) and (c) we perform the same visualisation for two of the ablation models (3 and 5). Considering Fig. \ref{fig:context_embeddings} (b) we see that the model fails to learn such a discriminative embedding space. Furthermore, we would like to point out that in the proposed model the same input embedding is used to predict the future audio signal as well. The clear separation of the two classes (i.e speech and non-speech) verifies our hypothesis that jointly predicting the future audio signal for the next time window can improve SAD performance. To further demonstrate this ability in Fig \ref{fig:context_embeddings} (c) we visualise the input embeddings learnt through Ablation model 5 for the same set of examples, where the GAN based model only predicts the classification sequence without modelling the future audio signal. It is clear that the automatic loss function learning process has contributed to learning discriminative embeddings but we observe some areas with overlaps between the speech and non-speech embeddings in contrast to the clear segregation in Fig. \ref{fig:context_embeddings} (a). This further emphasises the importance of the joint learning of both tasks to better capture the discriminative features. 

\subsection{Time Complexity}

In order to demonstrate that the proposed TA-GAN approach is suitable for real-time use, we benchmarked the time complexity of TA-GAN on the test set of AMI Meeting corpus dataset on a single core of an Intel Xeon E5-2680 2.50GHz CPU and the TA-GAN model runs at 5.35 $\times$ faster than real time. The proposed system was able to generate 100 predictions (i.e,  100 seconds of audio) where the output is both 100 $\times$ 1 second length classification sequence predictions and 100 $\times$ 1 second length future audio predictions, in 18.70 seconds. In a similar setting both basic RNN and CG-­LSTM methods of \cite{gelly2018optimization} take approximately 8.56 seconds to generate 100 $\times$ 1 second length classification sequence predictions. It should be noted that the proposed TA-GAN model has a larger time complexity due to the joint prediction of both future audio and classification sequences. In terms of number of trainable parameters the proposed method contains 48K trainable parameters while the basic RNN and CG-­LSTM methods of \cite{gelly2018optimization} have 6K trainable parameters. 

\section{Conclusion}
In this paper, we propose a novel multi-task learning framework for speech activity detection, by properly analysing the context of the input embeddings and their temporal accordance. We contribute a novel data-driven method to capture salient information from the observed audio segment by jointly predicting the speech activity classification sequence and the audio for the next time frame. Additionally, we introduce a temporal discriminator to enforce these relationships in the synthesised data. 

Our quantitative evaluations using multiple supervised SAD benchmarks, including NIST OpenSAT' 17 \cite{nist_open_SAT}, AMI Meeting \cite{carletta2006announcing} OpenKWS'13 \cite{harper2014iarpa} and HAVIC \cite{havic_ldc} demonstrated the utility of the proposed multi-task learning framework compared to the single task based supervised SAD baselines. Furthermore, through ablation model evaluations presented Sec. \ref{sec:ablation} we demonstrate that the automatic learning of a loss function specifically considering the task at hand, as opposed to using hand engineered losses, has significantly contributed to the superior performance attained in the proposed multi-task learning framework. 

In addition, in Tab. \ref{tab:ablation} we provide comparisons regarding systems with and without using the proposed temporal discriminator. The evaluation of the temporal discriminator, which enforces the temporal relationships between audio frames of the synthesised outputs, demonstrates the utility of incorporating this intelligence in the discriminator, which guides the generator to generate realistic outputs. With empirical evaluations we illustrate that the future audio segment prediction auxiliary task contributes to augment the performance of the SAD task, demonstrating the utility of multi-task learning and the importance of the future audio segment prediction task for learning the context of the input embeddings. 

To better demonstrate the robustness of the proposed framework we conducted a cross-database evaluation where we train the model using a seperate dataset and tested on another dataset. This experiment revealed that the proposed multi-task learning framework learns better discriminative features which are more robust across multiple datasets, compared to the current state-of-the-art supervised SAD models. We would like to emphasise that these evaluated datasets are of different languages, accents, and acoustics and the proposed method exhibits 37-52\% relative gain over the best alternate approach (ACAM \cite{kim2018voice}) when evaluated with NIST OpenSAT' 17 \cite{nist_open_SAT}.


\section*{Acknowledgment}
This research was supported by an Australian Research Council (ARC) Discovery grant DP140100793.

\bibliographystyle{IEEEtran}
\bibliography{refs.bib}

\begin{IEEEbiography}[{\includegraphics[width=1in,height=1.25in,clip,keepaspectratio]{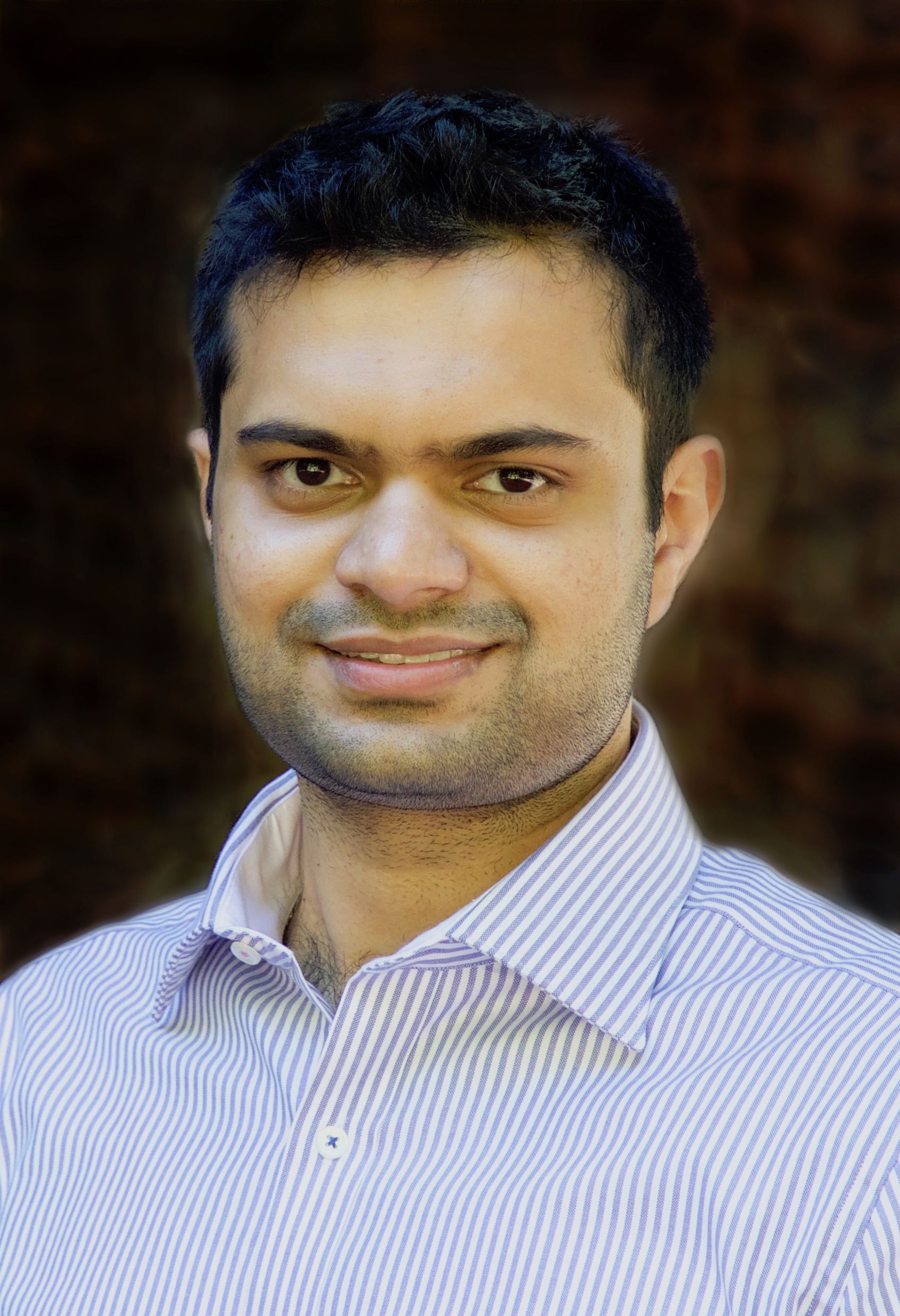}}]{Tharindu Fernando} received his BSc (special degree in computer science) from the University of Peradeniya, Sri Lanka and his PhD from Queensland University of Technology (QUT), Australia, respectively. He is currently a Postdoctoral Research Fellow in the SAIVT Research Program of School Electrical Engineering and Computer Science at QUT. His research interests focus mainly on human behaviour analysis and prediction. 
\end{IEEEbiography}

\begin{IEEEbiography}[{\includegraphics[width=1in,height=1.25in,clip,keepaspectratio]{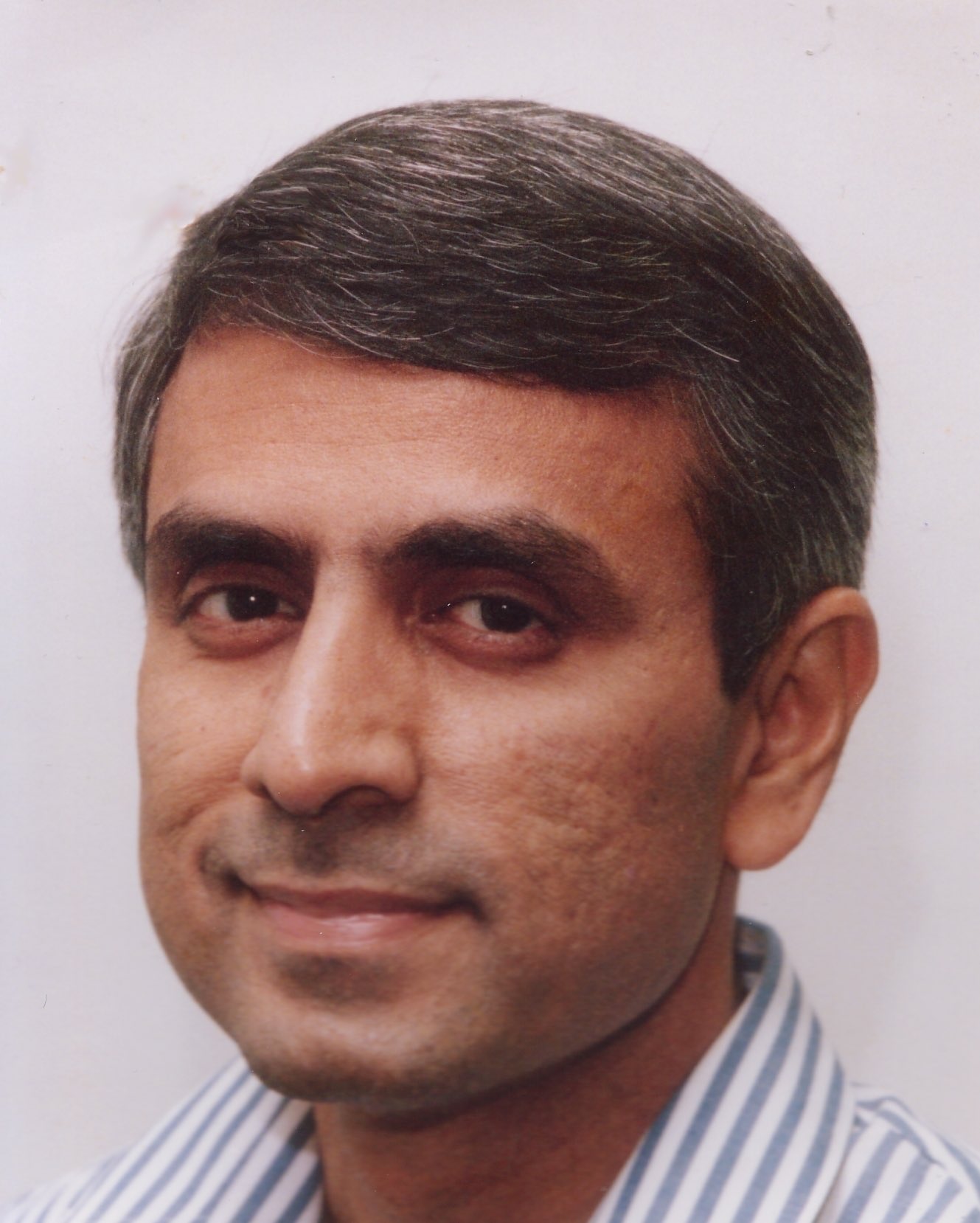}}]{Sridha Sridharan} has a BSc (Electrical Engineering) degree and obtained a MSc (Communication Engineering) degree from the University of Manchester, UK and a PhD degree from University of New South Wales, Australia. He is currently with the Queensland University of Technology (QUT) where he is a Professor in the School Electrical Engineering and Computer Science. Professor Sridharan is the Leader of the Research Program in Speech, Audio, Image and Video Technologies (SAIVT) at QUT, with strong focus in the areas of computer vision, pattern recognition and machine learning. He has published over 600 papers consisting of publications in journals and in refereed international conferences in the areas of Image and Speech technologies during the period 1990-2019. During this period he has also graduated 75 PhD students in the areas of Image and Speech technologies. Prof Sridharan has also received a number of research grants from various funding bodies including Commonwealth competitive funding schemes such as the Australian Research Council (ARC) and the National Security Science and Technology (NSST) unit. Several of his research outcomes have been commercialised.
\end{IEEEbiography}

\begin{IEEEbiography}[{\includegraphics[width=1in,height=1.25in,clip,keepaspectratio]{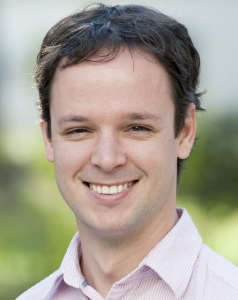}}]{Mitchell McLaren}, Ph.D., is a senior computer scientist in SRI International's Speech Technology and Research (STAR) Laboratory. His research interests include speaker and language identification, as well as other biometrics such as face recognition. Prior to joining SRI in 2012, Mitchell was a postdoctoral researcher and the University of Nijmegen, The Netherlands where he focused on speaker and face identification on the Bayesian Biometrics for Forensics (BBfor2) project, funded by Marie Curie Action. His Ph.D. in speaker identification is from the Queensland University of Technology (QUT), Brisbane, Australia.
\end{IEEEbiography}

\begin{IEEEbiography}[{\includegraphics[width=1 in,height=1.25in,clip,keepaspectratio]{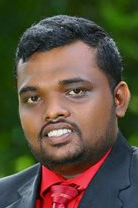}}]{Darshana Priyasad} is a PhD student at Queensland University of Technology, Australia. He received his Bachelor of Science in Engineering, specialised in Integrated Computer Engineering with first class honours from the University of Moratuwa, Sri Lanka. His research interests include deep learning, computer and machine vision.
\end{IEEEbiography}

\begin{IEEEbiography}[{\includegraphics[width=1in,height=1.25in,clip,keepaspectratio]{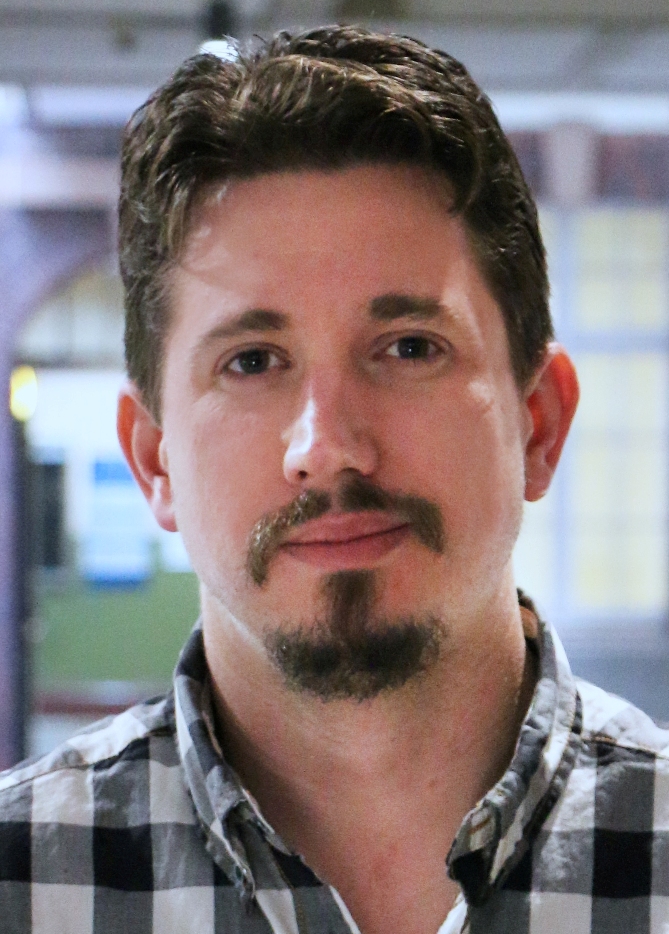}}]{Simon Denman} received a BEng (Electrical), BIT, and PhD in the area of object tracking from the Queensland University of Technology (QUT) in Brisbane, Australia. He is currently a Senior Research Fellow with the Speech, Audio, Image and Video Technology Laboratory at QUT. His active areas of research include intelligent surveillance, video analytics, and video-based recognition.
\end{IEEEbiography}

\begin{IEEEbiography}[{\includegraphics[width=1in,height=1.25in,clip,keepaspectratio]{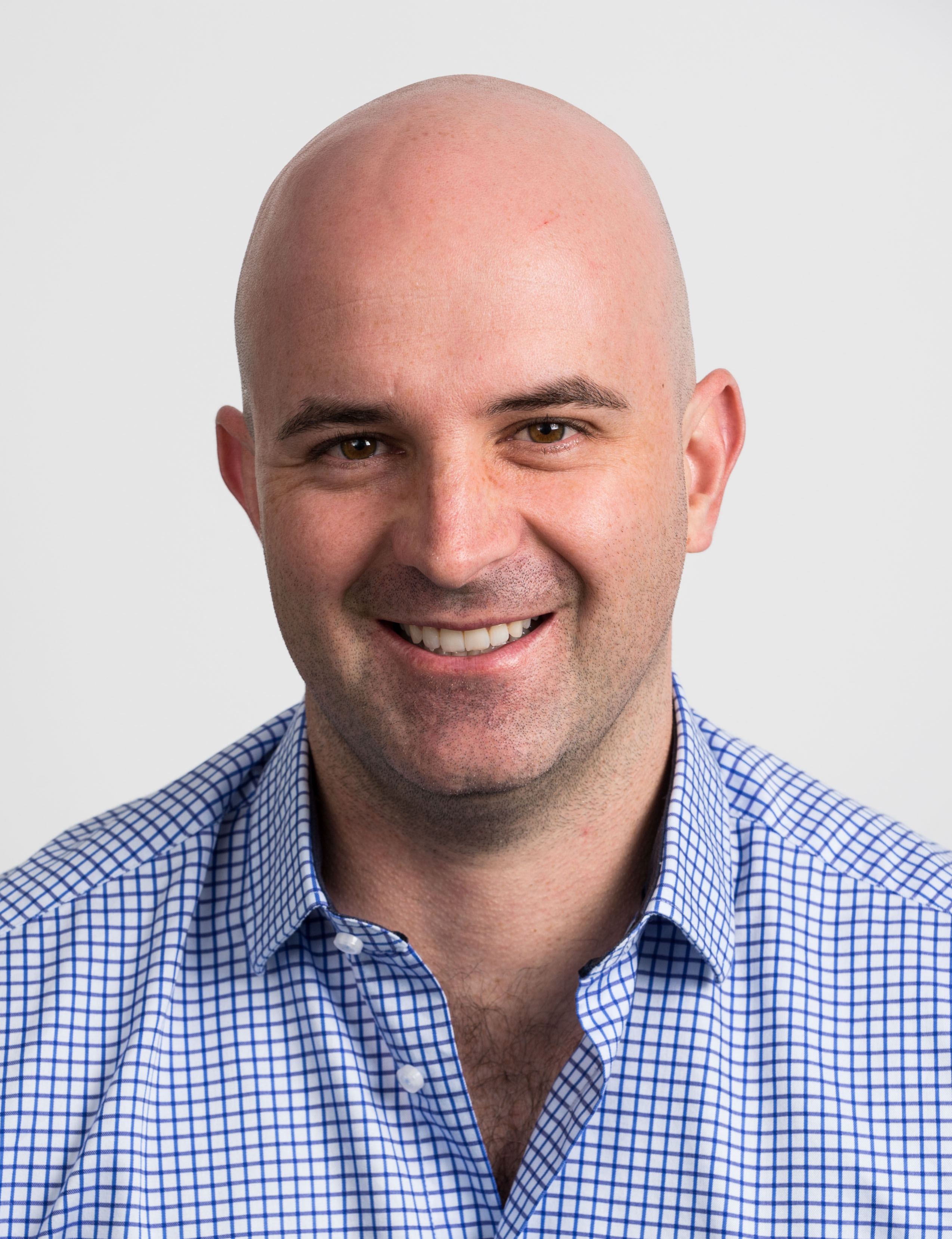}}]{Clinton Fookes} (SM'06) received his B.Eng. (Aerospace/Avionics), MBA, and Ph.D. degrees from the Queensland University of Technology (QUT), Australia. He is currently a Professor and Head of Discipline for Vision and Signal Processing within the Science and Engineering Faculty at QUT. He actively researchers across computer vision, machine learning, and pattern recognition areas. He serves on the editorial board for the IEEE Transactions on Information Forensics \& Security. He is a Senior Member of the IEEE, an Australian Institute of Policy and Science Young Tall Poppy, an Australian Museum Eureka Prize winner, and a Senior Fulbright Scholar.
\end{IEEEbiography}

\end{document}